\documentstyle[12pt]{article}

\begin{document}


\begin{center}
GASDYNAMIC MODEL OF STREET CANYON: PCCMM'99 POSTER SESSION
\end{center}
\begin{center}
Maciej M. Duras
\end{center}
\begin{center}
Institute of Physics, Cracow University of Technology, 
ulica Podchor\c{a}\.zych 1, 30-084 Cracow, Poland
\end{center}
\begin{center}
Email: mduras @ riad.usk.pk.edu.pl
\end{center}
\begin{center}
"The XIV Polish Conference on Computer Methods in Mechanics PCCMM99"; 
26th May 1999 - 28th May 1999; 
Rzesz\'{o}w, Poland (1999). Poster session.
\end{center}

\section{Abstract}
A general proecological urban road traffic control idea
for the street canyon is proposed
with emphasis placed on development of advanced
continuum field gasdynamical (hydrodynamical) 
control model of the street canyon.  
The continuum field model of optimal control of street canyon is studied.
The mathematical physics approach (Eulerian approach) 
to vehicular movement,
to pollutants' emission and to pollutants' dynamics is used.
The rigorous mathematical model is presented,
using gasdynamical (hydrodynamical) theory for both air constituents
and vehicles, including many types of vehicles and many types
of pollutant (exhaust gases) emitted from vehicles.
The six optimal control problems are formulated
and numerical simulations are performed.
Comparison with measurements are provided.
General traffic engineering conclusions are inferred.
[PCCMM'99 poster session].

\section{Description of the model}
The vehicular flow is multilane bidirectional one-level
one-dimensional rectilinear and it is considered
with two coordinated signalized junctions
\cite{Duras 1998 thesis, Duras 1996 Polish}.
The vehicles belong to different vehicular classes:
passenger cars, trucks.
The emission from vehicles are based on technical measurements
and many types of pollutants are considered
(carbon monoxide CO, hydrocarbons HC, nitrogen oxides NOx).
The vehicular dynamics is based on hydrodynamical approach
\cite{Michalopoulos 1984}.
The governing equations are continuity equation
for number of vehicles 
and Greenshields' equilibrium speed-density u-k model
\cite{Greenshields 1934}.
The model of dynamics of pollutants is also hydrodynamical.
The model consists of a set of mutually interconnected
vector nonlinear, spatially three-dimensional, temporally dependent, 
partial differential equations with nonzero righthand sides (sources),
and of boundary and initial problem.
The pollutants, oxygen, and remaining gaseous constituents of air,
are treated as mixture of noninteracting, Newtonian, viscous
fluid (perfect or ideal gases).
The model incorporates as variables the following fields:
density of mixture, mass concentrations of constituents of mixture,
velocity of mixture, temperature of mixture, pressure of mixture,
and intrinsic (internal) energy of mixture.
The model is based on assumption 
on local laws of balance (of conservation): 
of mass of mixture, of masses of constituents of mixture,
of momentum of mixture, of energy of mixture,
and of state equation (Clapeyron's law).
The model of dynamics is solved by finite difference
scheme.

The six optimization problems are formulated
by defining functionals of total travel time,
of global emissions of pollutants, 
and of global concentrations of pollutants,
both in the studied street canyon,
and in its two nearest neighbour substitute canyons.
The vector of control is a five-tuple
composed of two cycle times, two green times,
and one offset time between the traffic lights.
The optimal control problem consists of
minimization of the six functionals
over admissible control domain.

\section{Equations of dynamics.}

{\bf E1. Balance of momentum - Navier Stokes equation.}

\begin{eqnarray}
& & \rho (\frac{\partial \vec{v}}{\partial t}
+ (\vec{v} \circ \nabla) \vec{v}) + S\vec{v}=
\label{Navier-Stokes-eq} \\
& & = - \nabla p + \eta \Delta \vec{v} 
+ (\xi + \frac{\eta}{3}) \nabla ({\rm div} \vec{v}) + \vec{F}.
\nonumber
\end{eqnarray}

{\bf E2. Balance of mass - Equation of continuity.}
\begin{equation}
\frac{\partial \rho}{\partial t}
+ {\rm div}(\rho \vec{v})=S.
\label{continuity-eq}
\end{equation}

{\bf E3. Balance of mass of constituents - Diffusion equations.}

\begin{eqnarray}
& & \rho (\frac{\partial c_{i}}{\partial t} + \vec{v} \circ \nabla c_{i})=
Set_{i} - c_{i} S + 
\label{diffusion-eq-a} \\
& & + \sum_{m=1}^{N-1}\{ (D_{im}-D_{iN}) \cdot
\nonumber  \\
& & \cdot {\rm div}[\rho \nabla (c_{m}+\frac{k_{T, m}}{T} \nabla T)] \},
i=1, ..., N_{E}. \nonumber
\end{eqnarray}

\begin{eqnarray}
& & \rho (\frac{\partial c_{i}}{\partial t} + \vec{v} \circ \nabla c_{i})=
- c_{i} S +
\label{diffusion-eq-b} \\
& & + \sum_{m=1}^{N-1}\{ (D_{im}-D_{iN}) \cdot
\nonumber \\
& & \cdot {\rm div}[\rho \nabla (c_{m}+\frac{k_{T, m}}{T} \nabla T)] \},
\nonumber \\
& & i=(N_{E}+1), ..., N. \nonumber
\end{eqnarray}

{\bf E4. Balance of energy.}
\begin{eqnarray}
& &\rho (\frac{\partial \epsilon}{\partial t} + \vec{v} \circ \nabla \epsilon)=
\label{energy-eq} \\
& & =-(-\frac{1}{2}\vec{v}^{2} + \epsilon) S
+ \hat{{\cal T}}:\nabla \vec{v}+{\rm div}(-\vec{q})+\sigma.
\nonumber 
\end{eqnarray}

{\bf E5. Equation of state - Constitutive equation- Clapeyron's equation.}
\begin{equation}
\frac{p}{\rho}=\frac{R}{m_{air}}T,
\label{state-eq}
\end{equation}

\begin{equation}
p_{i}=c_{i}\frac{m_{air}}{m_{i}}p,
i=1, ..., N.
\label{constituent-state-eq}
\end{equation}

{\bf E6. Balance of vehicles - Equation of continuity.}
\begin{equation}
\frac{\partial k_{l, vt}^{{\rm L}}}{\partial t} 
+ {\rm div}(k_{l, vt}^{{\rm L}}\vec{w}_{l, vt}^{{\rm L}})=0.  
\label{vehicle-continuity-eq-a}
\end{equation}  

\begin{equation}
\frac{\partial k_{r, vt}^{{\rm R}}}{\partial t} 
+ {\rm div}(k_{r, vt}^{{\rm R}}\vec{w}_{r, vt}^{{\rm R}})=0.  
\label{vehicle-continuity-eq-b}
\end{equation}   

\[
vt=1, ..., VT, l=1, ..., n_{{\rm L}},
r=1, ..., n_{{\rm R}}.  
\]

\section{\bf Optimization problems.}

{\bf E8. Control.}
\begin{equation}
u=(g_{1}, C_{1}, g_{2}, C_{2}, F) \in U^{{\rm adm}},
\label{control-5-tuple-eq}
\end{equation}
where $g_{i}$, are green times, 
$C_{i}$, are cycle times,
$F$ is offset time, 
and $U^{{\rm adm}}$ is a set of admissible control variables.

{\bf F1. Total travel time for single canyon.} 
\begin{equation}
J_{{\rm TTT}}^{*}=J_{{\rm TTT}}(u^{*})=\inf \{ u \in U^{{\rm adm}}: J_{{\rm TTT}}(u) \},
\label{optimum-TTT-def}
\end{equation}   
  
\begin{eqnarray}
& & J_{{\rm TTT}}(u)=
\label{functional-TTT-def} \\
& & =\sum_{l=1}^{n_{{\rm L}}} \sum_{vt=1}^{VT}
\int_{0}^{a} \int_{0}^{T}
k_{l, vt}^{{\rm L}}(x, t) dx \, dt
+\nonumber \\
& & + \sum_{r=1}^{n_{{\rm R}}} \sum_{vt=1}^{VT}
\int_{0}^{a} \int_{0}^{T}
k_{r, vt}^{{\rm R}}(x, t) dx \, dt.
\nonumber
\end{eqnarray}

{\bf F2. Global emission for single canyon.}
\begin{equation}
J_{{\rm E}}^{*}=J_{{\rm E}}(u^{*})=\inf \{ u \in U^{{\rm adm}}: J_{{\rm E}}(u) \},
\label{optimum-E-def}
\end{equation}   

\begin{eqnarray}
& & J_{{\rm E}}(u)=
\label{functional-E-def} \\
& & =\sum_{l=1}^{n_{{\rm L}}} \sum_{ct=1}^{CT} \sum_{vt=1}^{VT}
\int_{0}^{a} \int_{0}^{T}
e_{l, ct, vt}^{{\rm L}}(x, t) dx \, dt +
\nonumber \\
& & +
\sum_{r=1}^{n_{{\rm R}}} \sum_{ct=1}^{CT} \sum_{vt=1}^{VT}
\int_{0}^{a} \int_{0}^{T}
e_{r, ct, vt}^{{\rm R}}(x, t) dx \, dt.
\nonumber
\end{eqnarray}

{\bf F3. Global pollutants' concentration for single canyon.}
\begin{equation}
J_{{\rm C}}^{*}=J_{{\rm C}}(u^{*})=\inf \{ u \in U^{{\rm adm}}: J_{{\rm C}}(u) \},
\label{optimum-C-def}
\end{equation}   

\begin{equation}
J_{{\rm C}}(u)=
\sum_{i=1}^{n_{E}} 
\rho_{{\rm STP}} \int_{0}^{a} \int_{0}^{b} \int_{0}^{c} \int_{0}^{T}
c_{i}(x, y, z, t) dx \, dy \, dz \, dt.
\label{functional-C-def}
\end{equation}

{\bf F4. Total travel time for canyon in street subnetwork.}
\begin{eqnarray}
& & J_{{\rm TTT, ext}}^{*}=J_{{\rm TTT, ext}}(u^{*})=
\label{optimum-TTT-ext-def} \\
& & =\inf \{ u \in U^{{\rm adm}}: J_{{\rm TTT, ext}}(u) \},
\nonumber
\end{eqnarray}   

\begin{eqnarray}
& & J_{{\rm TTT, ext}}(u)=
J_{{\rm TTT}}(u) + 
\label{functional-TTT-ext-def} \\
& & +
\alpha_{{\rm TTT, ext}}^{1}
(a \sum_{l=1}^{n_{{\rm L}}} \sum_{vt=1}^{VT} k_{l, vt, {\rm jam}}^{{\rm L}})
(C_{1}-g_{1}) 
\nonumber \\
& & 
+
\alpha_{{\rm TTT, ext}}^{2}
(a \sum_{r=1}^{n_{{\rm R}}} \sum_{vt=1}^{VT} k_{r, vt, {\rm jam}}^{{\rm R}})
(C_{2}-g_{2}).
\nonumber
\end{eqnarray}

{\bf F5. Global emission for canyon in street subnetwork.}
\begin{equation}
J_{{\rm E, ext}}^{*}=J_{{\rm E, ext}}(u^{*})=\inf \{ u \in U^{{\rm adm}}: J_{{\rm E, ext}}(u) \},
\label{optimum-E-ext-def}
\end{equation}   

\begin{eqnarray}
& & J_{{\rm E, ext}}(u)=
J_{{\rm E}}(u) +
\label{functional-E-ext-def} \\
& & +
\alpha_{{\rm E, ext}}^{1}
(a \sum_{l=1}^{n_{{\rm L}}} \sum_{ct=1}^{CT} \sum_{vt=1}^{VT} 
e_{l, ct, vt, {\rm jam}}^{{\rm L}})
(C_{1}-g_{1})
\nonumber \\
& & +
\alpha_{{\rm E, ext}}^{2}
(a \sum_{r=1}^{n_{{\rm R}}} \sum_{ct=1}^{CT} \sum_{vt=1}^{VT} 
k_{r, ct, vt, {\rm jam}}^{{\rm R}})
(C_{2}-g_{2}).
\nonumber
\end{eqnarray}

{\bf F6. Global pollutants' concentration for canyon in street subnetwork.}
\begin{equation}
J_{{\rm C, ext}}^{*}=J_{{\rm C, ext}}(u^{*})=\inf \{ u \in U^{{\rm adm}}: J_{{\rm C, ext}}(u) \},
\label{optimum-C-ext-def}
\end{equation}   

\begin{eqnarray}
& & J_{{\rm C, ext}}(u)=
J_{{\rm C}}(u) +
\label{functional-C-ext-def} \\
& & +
(\rho_{{\rm STP}} a b c \sum_{l=1}^{n_{{\rm L}}} c_{i, STP})
\nonumber \\
& & 
[ 
\alpha_{{\rm C, ext}}^{1} (C_{1}-g_{1})
+
\alpha_{{\rm C, ext}}^{2} (C_{2}-g_{2})
]
.
\nonumber
\end{eqnarray}

\section{Conclusions.}

We assumed the data
from real street canyon Krasi\'nski Avenue 
in Cracow \cite{Brzezanski 1998}.
In general the numerical results
are in very good agreement 
with measured data from \cite{Brzezanski 1998}.

In general some vehicular traffic and hydrodynamical parameters
influence the solutions of optimization problems
{\bf F1, F2, F3, F4, F5, F6}, but not all of them:

{\bf R1.} The direction of velocity of air mixture is important.
The optimal pollutant concentrations for both single canyon
and canyon with its two substitute nearest neighbour canyons,
{\bf F3, F6},
are the lowest if the velocity components 
of the boundary and initial value problems
are equal to zero;
further, they are the grater 
if velocity has only nonzero vertical component $v_{z}$;
then, they are greater if velocity has only nonzero x-component $v_{x}$;
further, they are again greater if velocity has two nonzero components
$(0, v_{y}, v_{z})$;
next, they are again greater if velocity has three nonzero components
$(v_{x}, v_{y}, v_{z})$;
finally, they are the greatest if the velocity has only 
nonzero y-component $v_{y}$.

{\bf R2.} The optimal values for {\bf F1, F4} cases,
for {\bf F2, F5} cases, and for {\bf F3, F6} cases, 
decrease with ``uniformization'' of vehicles, 
when we pass from nonuniform vehicles  
to uniform ones. 
It is a result of decrement of the number of vehicles moving in the canyon. 
For uniform the values of maximum free flow speed, 
jam, saturation, threshold, green, and red densities take on minima.

{\bf R3.} The long vehicular queues decrease total travel times {\bf F1, F4},
and they increase both optimal emissions {\bf F2, F5},
and concentrations of pollutants {\bf F3, F6}.
The decrement of total travel times {\bf F1, F4},
with long vehicular queues is result of clustering of vehicles.

{\bf R4.} The constant of temperature scale  
does not differentiate the values 
of optimal concentrations of pollutants {\bf F3, F6}
in the temperature range near standard temperature and pressure STP 
conditions. However, it diminishes them even hundredfold
for very high temperatures.

{\bf R5.} The functional form of initial and boundary conditions
affects the optima. If they are constant then optimal concentrations
{\bf F3, F6} are twice higher than in the case when they
are changing exponentially in space in three dimensions. 

{\bf R6.} The presence of vehicles on both left and
right lanes is important.
The optimal total travel times and emissions are halved in absence of
vehicles on left or right lanes
with respect to situation when they circulate on both left and
right lanes.

{\bf R7.} The values of saturation, arrival, or jam vehicular density,
and of vehicular free flow velocities also affect the optima
{\bf F1, F2, F3, F4, F5, F6}.

{\bf R8.} The assumption of energy conservation equation,
of thermodiffusion effect,
of chemical potential and of Grand Canonical ensemble,
as well as of influence of gravity on intrinsic energy and
on chemical potential,
drastically changes the optimal concentrations {\bf F3, F6}
towards measured ones 
\cite{Duras 1996 Polish}. 

{\bf R9.} The value of time of simulation and of
discretization in time affects much the optima. 
The values of optimal solutions {\bf F1, F2, F3, F4, F5, F6}
increase from tenfold to hundredfold.
Also the optimal 5-tuples for {\bf F1, F2, F3, F4, F5, F6}
change their values.
It is result of cumulative effect of length of period of simulation $T$
on integral functionals {\bf F1, F2, F3, F4, F5, F6}.


\end{document}